\documentstyle[11pt,aaspp4,twoside,flushrt]{article}
\begin{document}                                                      
\noindent{\bf                                                         
HELIUM MIXING IN GLOBULAR CLUSTER STARS
}                                                                     
\begin{list}{}{\topsep 0in                                            
          \partopsep 2\baselineskip                                   
          \itemsep 0pt                                                
          \parsep \baselineskip                                       
          \leftmargin .53in                                           
          \listparindent 0in                                          
          \labelsep 0in                                               
          \labelwidth 0in}                                            
\item ~                                                               

Allen V. Sweigart

Laboratory for Astronomy and Solar Physics, Code 681 \\
NASA/Goddard Space Flight Center, Greenbelt, MD 20771
\end{list}                                                              

\vspace{2\baselineskip}                                                
\noindent{ABSTRACT:}                                                  
The observed abundance variations in globular cluster red giants
indicate that these stars may be mixing helium from the hydrogen
shell outward into the envelope, presumably as a result of
internal rotation.  We have investigated the implications of such
helium mixing for both the red-giant-branch (RGB) and
horizontal-branch (HB) phases by computing a number of
noncanonical evolutionary sequences for different assumed mixing
depths and mass loss rates.  We find that the helium-mixed models
evolve to higher luminosities during the RGB phase and
consequently lose more mass than their canonical counterparts.
This enhanced mass loss together with the higher envelope helium
abundances of the helium-mixed models produces a markedly bluer
and somewhat brighter HB morphology.  As a result, helium mixing
can mimic age as a 2nd parameter and can reduce the ages of the
metal-poor globular clusters derived from the luminosity
difference between the HB and the main sequence turnoff.  Helium
mixing might also lead to a larger RR Lyrae period shift and to a
steeper slope for the RR Lyrae luminosity - metallicity relation
if the mixing is more extensive at low metallicities, as
suggested by the observed abundance variations.  We discuss the
implications of helium mixing for a number of other topics
including the low gravities of the blue HB stars, the origin of
the extreme HB stars, and the evolutionary status of the sdO
stars.  A variety of observational tests are presented to test
this helium-mixing scenario.
\vspace{\baselineskip}                                                
\pagestyle{myheadings}                                                
\markboth{\hspace*{1.0in}{\rm                                         
Sweigart
}\hspace{\fill}}{{\rm                                                 
Helium mixing
}}                                                               
\section{Introduction}
     Red-giant stars in individual globular clusters show large
star-to-star variations in the abundances of C, N, O, Na and Al
which are not predicted by canonical stellar evolution theory.
These variations are often anticorrelated in a manner suggestive
of nuclear processing and frequently become more pronounced with
evolution up the red-giant branch (RGB), as shown, for instance,
by the decline in the C abundance along the giant branches of M92
and NGC 6397 (Bell, Dickens, \& Gustafsson 1979; Carbon et al.
1982; Briley et al. 1990) and by the presence of super O-poor
stars near the tip of the RGB in M13 (Kraft et al. 1993, 1997).
These observational results indicate that low-mass red giants are
capable of mixing nuclearly processed material from the vicinity
of the hydrogen shell out to the surface (Kraft 1994).  Although
the nature of the mixing process is not well understood, it is
widely suspected that internal rotation plays a crucial role,
perhaps via meridional circulation (Sweigart \& Mengel 1979;
Peterson 1983; Norris 1987; Kraft 1994).

     The observed variations of Al (e. g., Norris \& Da Costa
1995; Kraft et al. 1997) are particularly important because they
require the deepest penetration of the mixing currents.  Langer,
Hoffman, \& Sneden (1993) have shown that Al can be produced in
low-mass red giants via proton captures on Mg, but this seems to
occur only within the hydrogen shell (Langer \& Hoffman 1995;
Cavallo, Sweigart, \& Bell 1996, 1997).  {\it Thus any mixing
process which dredges up Al will also dredge up helium.}  Since
the helium abundance is one of the main parameters determining
the stellar structure, such mixing could potentially have a major
impact on the luminosities, effective temperatures and lifetimes
of the subsequent evolutionary phases, especially the
horizontal-branch (HB) phase (Langer \& Hoffman 1995; Sweigart
1997).

     We have computed a number of noncanonical evolutionary
sequences in order to investigate the consequences of mixing
helium from the hydrogen shell into the envelope of a globular
cluster red-giant star (Sweigart 1997).  We describe these RGB
sequences as well as the numerical algorithm for helium mixing in
the following section.  The main results are given in Sec. 3,
where we briefly explore the implications of helium mixing for
four topics, namely, the morphology of the HB and the 2nd
parameter effect, the low gravities of the blue HB (BHB) stars,
the origin of the extreme HB (EHB) stars, and the evolutionary
status of the helium-rich sdO stars.  A short summary is given in
Sec. 4.

\section{RGB Evolutionary Sequences with Helium Mixing}
     We first discuss the composition distribution around the
hydrogen shell of a typical RGB model in order to emphasize the
connection between the observed Al enhancements and helium
mixing.  Figure 1 gives a representative distribution taken from
the results of Cavallo et al. (1997).  Although the details of
the composition distribution in Figure 1 can vary somewhat from
case to case, the main features should be invariant.  In
particular, we note that there is a region above the hydrogen
shell within which C is depleted by the CN cycle and, somewhat
closer to the shell, a region within which O is depleted by the
ON cycle.  The dredge-up of CN and ON processed material from
these regions would alter the envelope CNO abundances even if the
mixing currents do not penetrate into the shell.  Thus the
observed CNO variations by themselves do not necessarily imply
helium mixing.   The same is also true for $^{23}{\rm Na}$ which
is enhanced above the shell by proton capture on $^{22}{\rm
Ne}$, although larger $^{23}{\rm Na}$ enhancements would require
mixing into the shell where proton capture on the more abundant
$^{20}{\rm Ne}$ can take place (Denisenkov \& Denisenkova
1990).  In contrast, $^{27}{\rm Al}$ is only produced within the
shell because of the higher temperatures needed for proton
capture by Mg.  It follows therefore that the dredge-up of Al, as
implied by the observations, will be accompanied by the dredge-up
of helium.  In some cases helium might be dredged up from the
upper part of the shell, where Al is not always enhanced.  Thus
an enhancement in the surface Al abundance would be a sufficient,
but not necessary, condition for helium mixing.

\begin{figure}[t]
  \epsscale{0.60}
  \plotone{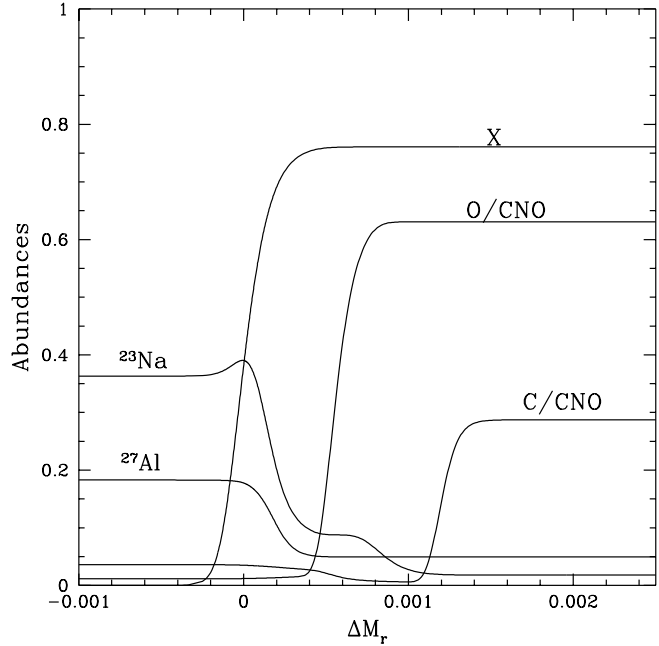}
\caption{ Composition distribution around the hydrogen shell of
a typical RGB model.  The hydrogen shell is labeled by the
hydrogen abundance X.  The C and O abundances are given as
fractions of the total CNO abundance, while the $^{23}{\rm Na}$
and $^{27}{\rm Al}$ abundances have been multiplied by factors
of 10 and 20, respectively.  The abscissa gives the difference in
the mass coordinate ${\rm M}_{\rm r}$ in solar units relative
to the center of the shell. }
\end{figure}

     Helium mixing was included in the model calculations with
the algorithm illustrated in Figure 2.  We assume that the mixing
currents are able to penetrate into the shell and consequently
that some of the helium produced by the hydrogen burning is added
to the envelope instead of the core.  The depth of the mixing is
specified by the parameter $\Delta {\rm X}_{\rm mix}$ which
measures the difference in the hydrogen abundance X between the
envelope (${\rm X} = {\rm X}_{\rm env}$) and the innermost
point reached by the mixing currents (${\rm X} = {\rm X}_{\rm
env} - \Delta {\rm X}_{\rm mix}$).  All of the helium produced
outside the point where ${\rm X} = {\rm X}_{\rm env} - \Delta
{\rm X}_{\rm mix}$ is mixed outward into the envelope where it
is diluted throughout the entire envelope mass.  As a result, the
envelope hydrogen abundance gradually decreases from one model to
the next.  All of the helium produced interior to the point where
${\rm X} = {\rm X}_{\rm env} - \Delta {\rm X}_{\rm mix}$ is
added to the core as in canonical models.  For simplicity we
assume that the penetration depth $\Delta {\rm X}_{\rm mix}$
is constant during the evolution up the RGB.  This assumption
will be relaxed in future computations.

\begin{figure}[t]
  \epsscale{0.75}
  \plotone{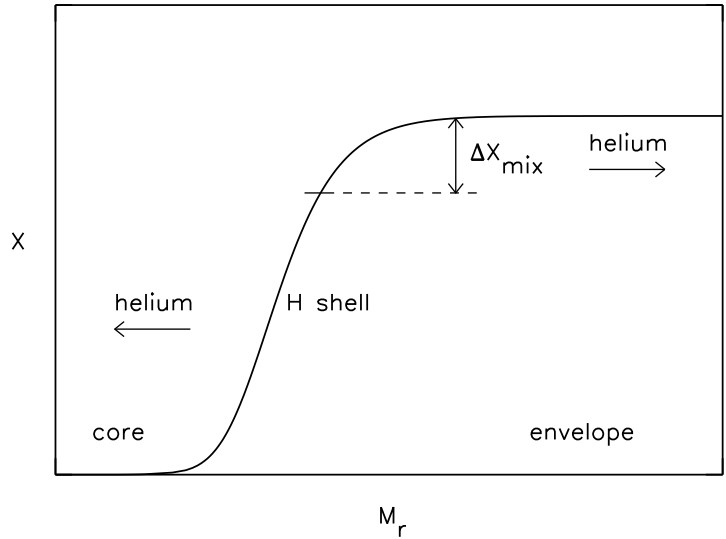}
\caption{ Schematic representation of the algorithm for helium
mixing in RGB models.  The solid curve gives the hydrogen
abundance X within the hydrogen shell as a function of the mass
coordinate ${\rm M}_{\rm r}$. }
\end{figure}

     Using this algorithm, we have evolved 53
sequences up the RGB to the helium flash for various amounts of
helium mixing and mass loss.  More specifically, we have computed
sequences for $\Delta {\rm X}_{\rm mix} \, = \, 0.0$ (no
helium mixing), 0.05, 0.10 and 0.20 and for $\eta_{\rm R} \, =
\, 0.0$ (no mass loss) to 0.8, where $\eta_{\rm R}$ is the
Reimers (1975) mass loss parameter introduced by Fusi Pecci \&
Renzini (1976).  Helium mixing was begun at the point along the
RGB where the hydrogen shell burns through the hydrogen
discontinuity produced by the deep penetration of the convective
envelope during the first dredge-up.  This point corresponds to
the well-known ``bump" in the RGB luminosity function (Fusi Pecci
et al. 1990 and references therein).  Each sequence had the same
mass and composition at the zero-age main sequence, i.e., mass M
= $0.805 \, \, {\rm M}_{\sun}$, helium abundance Y = 0.23 and
heavy-element abundance Z = 0.0005.  The input physics for these
sequences is outlined by Sweigart (1997).

     We emphasize that the results presented here are based on a
single value of Z and on the mixing algorithm in Figure 2.
Consequently they explore just a small part of parameter space.
We will extend these preliminary results in future calculations,
especially to higher metallicities.

\begin{figure}[t]
  \epsscale{0.75}
  \plotone{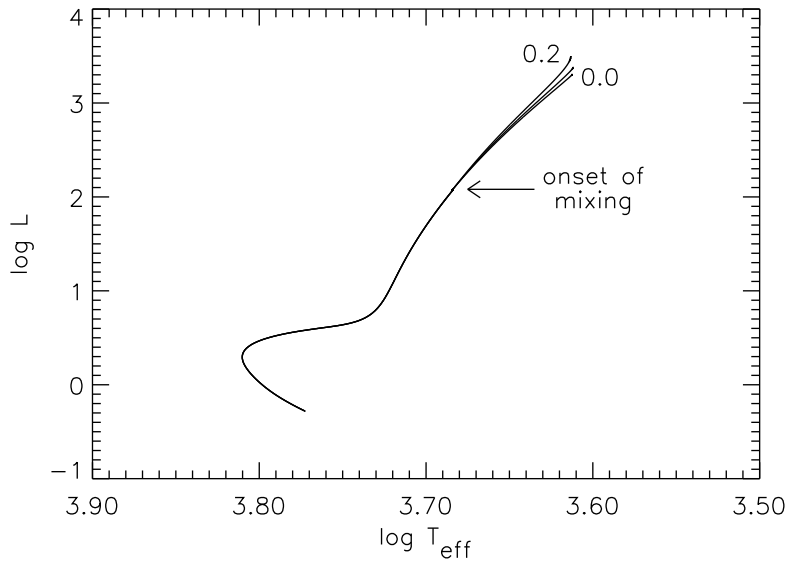}
\caption{ RGB evolutionary tracks for the Reimers mass loss
parameter $\eta_{\rm R} \, = \, 0.3$ and 3 depths of helium
mixing:  $\Delta {\rm X}_{\rm mix} \, = \, 0.0$ (canonical
evolution), 0.1 and 0.2. }
\end{figure}

     With this caveat we now consider the effects of helium
mixing on the RGB evolution.  Figure 3 shows 3 RGB tracks
computed for the mass loss parameter $\eta_{\rm R} \, = \,
0.3$: a canonical track labeled by $\Delta {\rm X}_{\rm mix}
\, = \, 0.0$ and 2 helium-mixed tracks for $\Delta {\rm
X}_{\rm mix} \, = \, 0.1$ and 0.2.  As indicated previously,
helium mixing begins at the bump in the RGB luminosity function
which in this case occurs at log L = 2.1.  Prior to this point
the tracks in Figure 3 are identical.  Helium mixing along the
$\Delta {\rm X}_{\rm mix} \, = \, 0.1$ and 0.2 tracks leads
to a progressive increase in the envelope helium abundance ${\rm
Y}_{\rm env}$ (see Figure 2 of Sweigart 1997).  By the tip of
the RGB, ${\rm Y}_{\rm env}$ in these sequences exceeds the
canonical value by 0.07 and 0.19, respectively.  Despite this
substantial increase in ${\rm Y}_{\rm env}$, the helium-mixed
and canonical tracks have quite similar morphologies.  In
particular, the helium-mixed tracks are shifted blueward by only
a few 0.01 mag in \bv, which would be difficult to detect,
given the paucity of stars near the tip of the RGB in the
globular clusters.

     One important consequence of helium mixing is an increase in
the RGB tip luminosity.  This occurs because the higher envelope
helium abundance in the helium-mixed models affects both the
hydrostatic structure around the hydrogen shell by increasing the
mean molecular weight and the thermal structure by decreasing the
opacity.  Both of these effects act together to increase the
hydrogen burning rate and hence the surface luminosity.  As a
result, the helium-mixed sequences lose significantly more mass
during the RGB phase for a given value of $\eta_{\rm R}$, as
illustrated in Figure 3 of Sweigart (1997).  This suggests that a
spread in the stellar rotation rate might lead to a spread in the
final RGB mass, should rotation control the amount of helium
mixing.

     The correlation between the Al enhancement and the envelope
helium abundance predicted by helium mixing implies that the 
Al-enhanced giants in a cluster such as M13 should appear to be more
metal-rich than the Al-normal giants.  The size of the predicted
spread in [Fe/H] is, however, quite small, amounting to only 0.04
and 0.12, respectively, by the tip of the RGB in the $\Delta
{\rm X}_{\rm mix} \, = \, 0.1$ and 0.2 tracks in Figure 3.
Although the detection of such a small spread in [Fe/H] would be
a formidable undertaking, there are some factors which would
reduce the observational uncertainties.  Most importantly, this
test depends on a differential, not absolute, measurement of
[Fe/H].  Moreover, one could compare mean [Fe/H] values for the
Al-enhanced and Al-normal giants.

\section{Implications for HB and Post-HB Evolution}
\subsection{Horizontal-Branch Morphology: 2nd Parameter Effect}
     In order to investigate the effects of helium mixing on the
post-RGB evolution, we have continued all of the RGB sequences
through the helium flash to the zero-age horizontal branch (ZAHB)
and then through the HB and post-HB phases until the onset of the
helium-shell flashes.  Using these evolutionary sequences, we
will now address four aspects of the HB and post-HB evolution,
beginning with the morphology of the HB and the 2nd parameter
effect.

     From the previous discussion we know that a helium-mixed
star will arrive on the ZAHB with a higher envelope helium
abundance and a lower mass than the corresponding canonical star
and hence will lie at a higher effective temperature.  In fact, a
star's ZAHB location turns out to be remarkably sensitive to the
amount of helium mixing.  We illustrate this point in Figure 4,
where we present 4 synthetic HB simulations (Catelan 1996) for
various amounts of helium mixing.  The canonical simulation given
in panel (a) assumes a Gaussian distribution of the HB mass
${\rm M}_{\rm HB}$ with a mean mass $\langle {\rm M}_{\rm
HB} \rangle$ of 0.675 ${\rm M}_{\sun}$, corresponding to a
mean mass loss parameter $\langle \eta_{\rm R} \rangle$ of
0.34, and a mass dispersion $\sigma_{\rm M}$ of 0.03 ${\rm
M}_{\sun}$.  These values were chosen to give an M3-like HB
morphology with approximately equal numbers of stars blueward and
redward of the instability strip, as indicated by the HB
parameters (cf. Buonanno 1993) listed in the upper left corner of
panel (a).  The simulations in the remaining panels of Figure 4
then adopt the same distribution in $\eta_{\rm R}$ as implied
by the mass distribution in panel (a).

\begin{figure}[t]                                                               
\figurenum{4}                                                                  
\plotfiddle{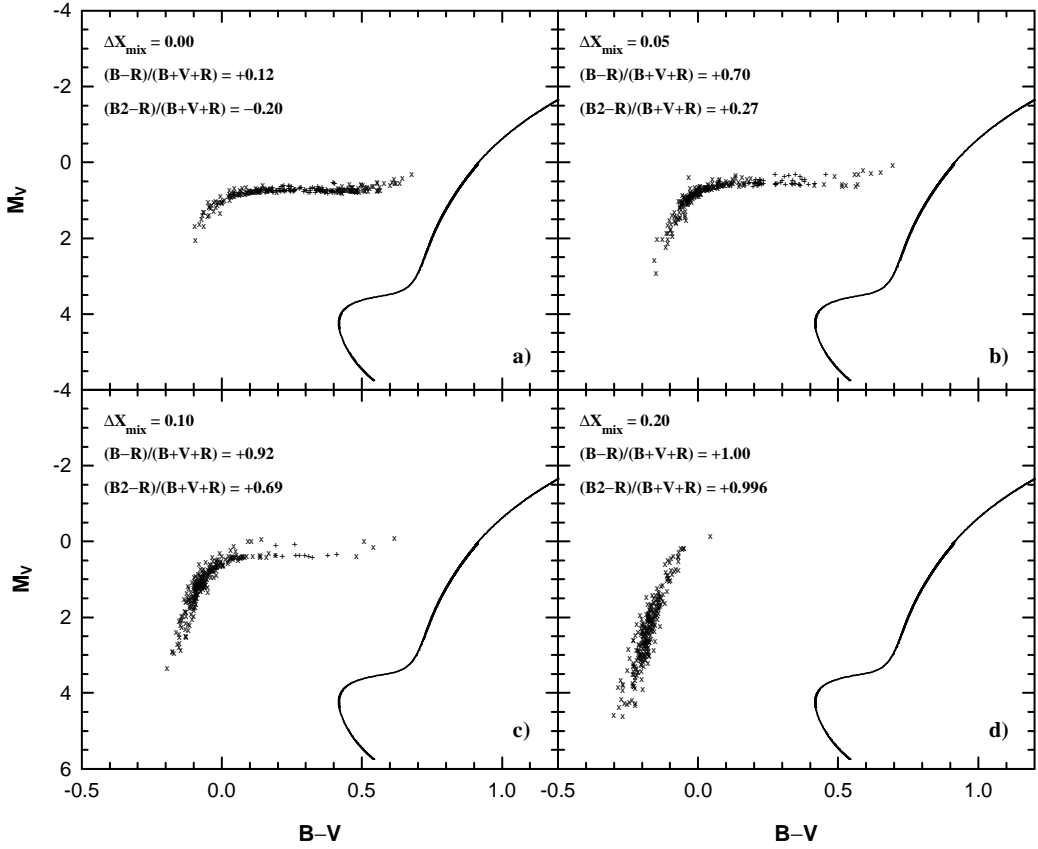}{4.45in}                                          
{0}{70}{70}{-215}{-197.0}                                                    
\caption{ Synthetic HB simulations for canonical evolution
(panel a) and for 3 depths of helium mixing: $\Delta {\rm
X}_{\rm mix} \, = \, 0.05$ (panel b), 0.10 (panel c) and 0.20
(panel d).  Plus signs denote RR Lyrae variables. }   
\end{figure}  

     The simulation in panel (b) for $\Delta {\rm X}_{\rm mix}
\, = \, 0.05$ shows that helium mixing leads to a markedly
bluer HB morphology even though the mean increase in ${\rm
Y}_{\rm env}$ in this case is only $\approx 0.03$.  Besides
being bluer, the HB in panel (b) is also brighter by $\approx
0.2$ mag, implying a corresponding increase in the luminosity
difference $\Delta {\rm V}$ between the HB and the main
sequence turnoff.  It follows that the ``apparent" age derived
from the $\Delta {\rm V}$ method (e.g., Iben \& Renzini 1984)
for the cluster in panel (b) would be 2 - 3 Gyr older than for
the canonical cluster in panel (a).  This is approximately the
age difference required to explain the change in the HB
morphology between panels (a) and (b), assuming that age is the
only parameter that varies between these clusters.  By shifting
the HB blueward and making the clusters appear older, helium
mixing could therefore mimic age as a 2nd parameter.  This would
obviously affect estimates of the formation timescale of the
galactic halo derived from the 2nd parameter clusters.

     Besides affecting the relative ages of the 2nd parameter
clusters, helium mixing could also affect the absolute globular
cluster ages, particularly for the most metal-poor (and
presumably oldest) clusters for which the observed abundance
variations in CNO (Bell \& Dickens 1980) and possibly Al (Norris
\& Da Costa 1995) are most pronounced.  The fact that the most
metal-poor globular clusters do not have the bluest HB's may be
an indication that the mass loss efficiency decreases at the
lowest metallicities (Renzini 1983; Fusi Pecci et al. 1993).  The
present sequences show that helium mixing increases the ZAHB
luminosity ${\rm L}_{\rm HB}$ near the instability strip at the
rate $\Delta \, {\rm log} \, \, {\rm L}_{\rm HB} \,
\approx \, {\rm 1.5} \, \Delta {\rm X}_{\rm mix}$.
Combining this result with the dependence of the globular cluster
age on the turnoff luminosity given, for example, by VandenBerg,
Bolte, \& Stetson (1996), we obtain the relationship

\begin{equation}
\Delta {\rm t} \, \approx \, -50 \, \Delta {\rm X}_{\rm
mix}
\end{equation}

\noindent between the change in the globular cluster age in Gyr, as
inferred from $\Delta {\rm V}$, and the helium mixing parameter
$\Delta {\rm X}_{\rm mix}$.  We see that even a modest
penetration of the mixing currents into the hydrogen shell during
the RGB phase would significantly reduce the globular cluster
age, as a consequence of the higher luminosity of the 
helium-mixed HB models.  This may be especially important for 
clusters where the determination of $\Delta {\rm V}$ requires an
extrapolation from the blue HB (see discussion by Buonanno,
Corsi, \& Fusi Pecci 1989).  The ages estimated from $\Delta
{\rm V}$ in such cases might be too large if mixing is
responsible for the blue HB morphology.  A possible example of
this effect is the cluster M10, whose very blue HB seems to be
unusually bright (Hurley, Richer, \& Fahlman 1989; Arribas,
Caputo, \& Martinez-Roger 1991).  Interestingly, the $\Delta
{\rm V}$ value for M10 is the second largest in the sample of 43
clusters studied by Chaboyer, Demarque, \& Sarajedini (1996) and
is exceeded only by the value for NGC 6752, another cluster with
a purely blue HB.

     Panels (c) and (d) in Figure 4 illustrate how larger amounts
of helium mixing can shift the HB even further to the blue.  In
the $\Delta {\rm X}_{\rm mix} \, = \, 0.10$ simulation of
panel (c) most of the HB stars lie blueward of the instability
strip, while in the $\Delta {\rm X}_{\rm mix} \, = \, 0.20$
simulation in panel (d) the HB consists of a blue tail with a
population of EHB stars at its hot end.  The mean increase in
${\rm Y}_{\rm env}$ for these 2 simulations was $\approx 0.07$
and $\approx 0.20$, respectively.  We emphasize again that the
simulations in Figure 4 were all computed for the same
distribution of the Reimers mass loss parameter $\eta_{\rm R}$.

     In producing Figure 4 we assumed that all of the stars in
each panel mix to the same depth $\Delta {\rm X}_{\rm mix}$
along the RGB.  This may not be the case in an actual cluster if
the mixing is controlled by the stellar rotation rate $\omega$.
One would expect the mixing currents in a more rapidly rotating
red-giant star to penetrate more deeply and hence to dredge up
more processed material to the surface.  This might lead to a
gradient in the extent of the mixing along the HB, with the
hotter HB stars being more heavily mixed, as indicated
schematically in Figure 5.  In this case the BHB stars would be
helium-rich compared to the RR Lyrae stars which, in turn, might
be helium-rich compared to the red HB stars.  Such rotationally
driven mixing might thus lead to an upward tilt of the
``horizontal" part of the HB with decreasing \bv, as has been
found in the metal-rich clusters NGC 6388 and NGC 6441 by Piotto
et al. (1997).

\begin{figure}[t]
  \figurenum{5}
  \epsscale{0.65}
  \plotone{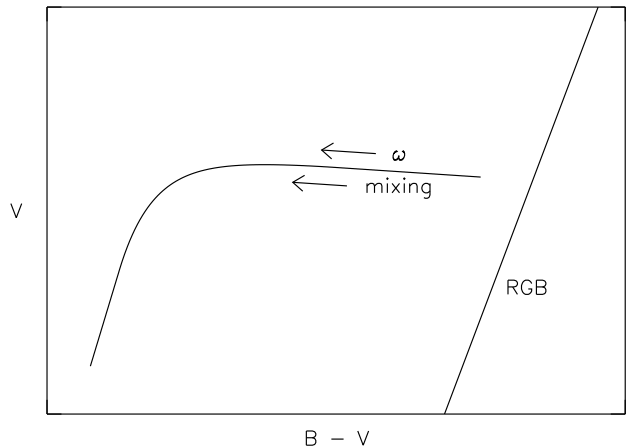}
\caption{ Schematic representation of the HB morphology produced
by variable stellar rotation rates $\omega$ and variable amounts
of helium mixing. }
\end{figure}

\begin{figure}[t]
  \figurenum{6}
  \epsscale{0.65}
  \plotone{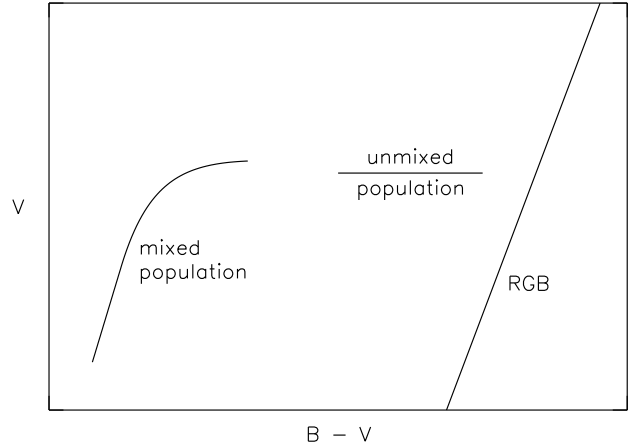}
\caption{ Schematic representation of the bimodal HB morphology
produced by a red unmixed population and a blue mixed population.
}
\end{figure}

     Figure 5 might be modified under some circumstances if there
should be a threshold in the stellar rotation rate for helium
mixing.  If the rotation rate is too slow, the mixing currents
might not be able to penetrate into the hydrogen shell before the
nuclear burning sets up a gradient in the mean molecular weight
$\mu$ large enough to inhibit the mixing.  This is more likely
to occur at high metallicities where the hydrogen shell is
steeper and where the timescale for setting up a $\mu$ gradient
is therefore shorter (Cavallo et al. 1997).  Thus helium mixing
in a metal-rich population might be confined to the more rapidly
rotating giants.  Such a threshold effect might lead to a bimodal
HB morphology consisting of a red unmixed population and a blue
mixed population, as shown schematically in Figure 6.  This
possibility may be relevant for understanding the bimodal HB
morphologies of some globular clusters (Crocker, Rood, \&
O'Connell 1988; Borissova et al. 1997) as well as the more
extreme bimodal morphologies found in metal-rich open clusters
such as NGC 6791 (Liebert, Saffer, \& Green 1994) and in
elliptical galaxies where a hot HB population is believed to
produce the UV upturn (Greggio \& Renzini 1990; Dorman,
O'Connell, \& Rood 1995).

     The above results support the possibility that stellar
rotation might control the extent of mixing along the RGB as well
as the well-known difference in HB morphology between some 2nd
parameter pairs of globular clusters (Renzini 1977; Fusi Pecci
\& Renzini 1978; Norris 1983, 1987; Peterson 1983; Buonanno,
Corsi, \& Fusi Pecci 1985; Kraft 1994; Kraft et al. 1995;
Peterson, Rood, \& Crocker 1995).  Under this scenario globular
clusters with rapid stellar rotation rates would mix helium along
the RGB and hence have a bluer than expected HB morphology.
Conversely, globular clusters with slower stellar rotation rates
would mix little, if any, helium, and hence their HB morphology
would approximate the canonical one.  An example of this
possibility is the 2nd parameter clusters M3 and M13, which some
authors suggest may have similar ages (Paltrinieri et al. 1997
and references therein).  Rotation may also be involved in
producing the correlation between the relative number of BHB
stars and the cluster central density (Fusi Pecci et al. 1993;
Buonanno et al. 1997).  As discussed by Buonanno et al. (1985),
stellar encounters within the dense cores of such globular
clusters might lead to a spin up of the red-giant envelope and
hence to enhanced mass loss (see also Renzini 1983).

     Helium mixing may also affect the properties of the RR Lyrae
variables by increasing their luminosity and hence their
predicted pulsation period P.  This may be relevant for
understanding the Sandage period-shift effect (Sandage 1982,
1990, 1993a).  For example, a period shift of $\Delta \, {\rm
log} \, {\rm P} \, = \, 0.065$ between M3 and M15 could be
explained if the red-giant stars in M15 mix into the depth
$\Delta {\rm X}_{\rm mix} \, = \, 0.06$ (Sweigart 1997).
Such mixing would have 2 important implications.  First, it would
increase ${\rm Y}_{\rm env}$ in the RR Lyrae variables in M15
by 0.04 and consequently produce a Y - Z anticorrelation between
these 2 clusters, as previously suggested by Sandage, Katem, \&
Sandage (1981) and Sandage (1982, 1990).  Although such an
anticorrelation between helium abundance and metallicity is
counterintuitive, it might arise during the RGB evolution if
metal-poor giants mix more deeply (VandenBerg \& Smith 1988).
The second implication would be an increase in the luminosity of
the RR Lyrae variables in M15 by $\approx 0.2$ mag and consequently 
a steeper slope for the RR Lyrae luminosity-metallicity 
relation, as has also been advocated by Sandage
(1982, 1990, 1993b).  We note that relatively bright luminosities
have been reported for the RR Lyrae variables in M15 by Bingham
et al. (1984), Simon \& Clement (1993) and Silbermann \& Smith
(1995).  From a Baade-Wesselink analysis, Storm, Carney, \&
Latham (1994) have found that the RR Lyrae variables in another
metal-poor cluster, M92, are $\approx 0.2$ mag brighter than the
field RR Lyrae variables of similar metallicity, although in this
case the effect could be partially due to evolution away from the
ZAHB (e.g., Rood \& Crocker 1989).  These results provide a
potential means for testing the helium-mixed models.  Before
doing so, however, one needs to understand why the Baade-Wesselink 
luminosities for field RR Lyrae variables are systematically 
fainter than canonical HB models (e.g., Castellani
\& De Santis 1994) and to determine whether this systematic
offset also applies to the cluster variables.  In any case, the
present results suggest that there may be a relationship between
the observed abundance variations along the RGB and the pulsation
properties of the RR Lyrae variables, as has also been noted by
VandenBerg et al. (1996).

\subsection{Low Gravities of the Blue HB Stars}
     A number of observational studies have shown that the
surface gravities of the globular cluster BHB stars are lower
than predicted by canonical stellar models (Crocker et al. 1988;
Moehler, Heber, \& de Boer 1995; de Boer, Schmidt, \& Heber
1995; Moehler, Heber, \& Rupprecht 1997b).  The size of this
discrepancy increases with ${\rm log} \, {\rm T}_{\rm eff}$,
reaching $\approx 0.3$ in $\Delta \, {\rm log} \, \, {\rm
g}$ for $4.3 \, > \, {\rm log} \, {\rm T}_{\rm eff} \, >
\, 4.2$.  Curiously the EHB stars with ${\rm log} \, {\rm
T}_{\rm eff} \, > \,  4.3$ have gravities in good agreement
with canonical theory.  These results are illustrated in Figure
7, where we plot the observed gravities for the hot HB stars in a
number of globular clusters together with a set of canonical HB
and post-HB evolutionary tracks.  Each track is represented by a
series of dots spaced every $5 \, \times \, {\rm 10}^{\rm
5}$ yr along the evolution so that one can better compare the
expected location of the models with the observational data.

     Figure 7 raises some intriguing questions: does the
structure of an HB star change in some basic way around ${\rm
log} \, {\rm T}_{\rm eff} \, = \, 4.3$, and could such a
change be related to the low observed gravities?  Although
perhaps a coincidence, ${\rm log} \, {\rm T}_{\rm eff} \, =
\, 4.3$ is approximately the temperature at which the hydrogen
shell becomes an active energy source.  Blueward of ${\rm log}
\, {\rm T}_{\rm eff} \, = \, 4.3$, the envelope mass is
insufficient to sustain an active hydrogen shell, while redward
of this temperature the hydrogen shell makes an increasingly
important contribution to the surface luminosity.  The observed
shift towards lower gravities at ${\rm log} \, {\rm T}_{\rm
eff} \, = \, 4.3$ may be an indication that the hydrogen shell
turns on more strongly as one goes from the EHB to the BHB than
predicted by canonical models.  Some support for this suggestion
is provided by the ultraviolet observations of the hot HB stars
in M79 (Hill et al. 1992), $\omega$ Cen (Whitney et al. 1994),
and NGC 6752 (Landsman et al. 1996b) obtained with the
Ultraviolet Imaging Telescope (Stecher et al. 1992).  In each of
these clusters the luminosity difference between the EHB and the
BHB appears to be larger than expected from canonical models.

\begin{figure}[t]
  \figurenum{7}
  \epsscale{0.75}
  \plotone{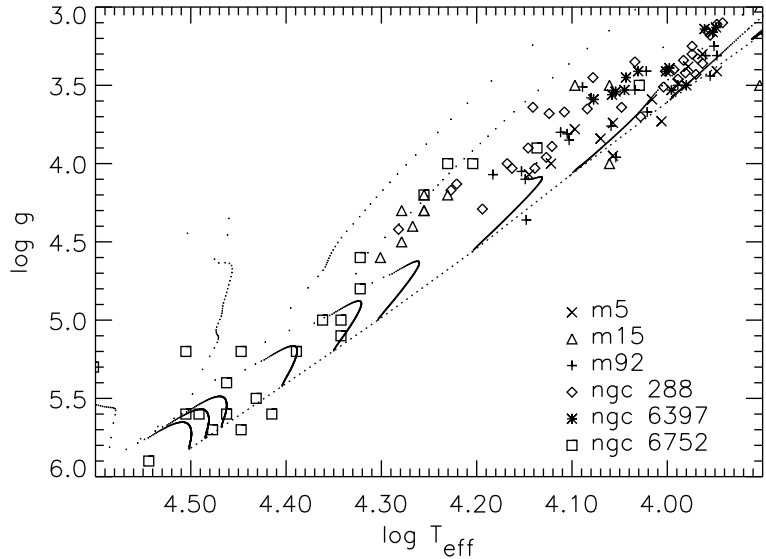}
\caption{ Comparison between canonical HB and post-HB
evolutionary tracks and the observed gravities of the hot HB
stars in a number of globular clusters.  The dots along each
track are separated by $5 \, \times \, {\rm 10}^{\rm 5}$ yr.
Data are from Crocker et al. (1988), Moehler et al. (1995, 1997b)
and de Boer et al. (1995). The dotted line indicates the ZAHB.}
\end{figure}

     Crocker et al. (1988) and Rood \& Crocker (1989) have
discussed the effects of various HB parameters on the predicted
gravities.  The only change they could identify which would
reproduce the observed gravities on both the EHB and BHB was an
enhancement in the envelope helium abundance (see also Gross
1973).  Changes in the CNO abundance, mass loss and age do not
work.  Similarly rotation is not a solution if its only effect is
to increase the core mass at the helium flash.

\begin{figure}[t]                                                               
  \figurenum{8}
  \epsscale{0.75}
  \plotone{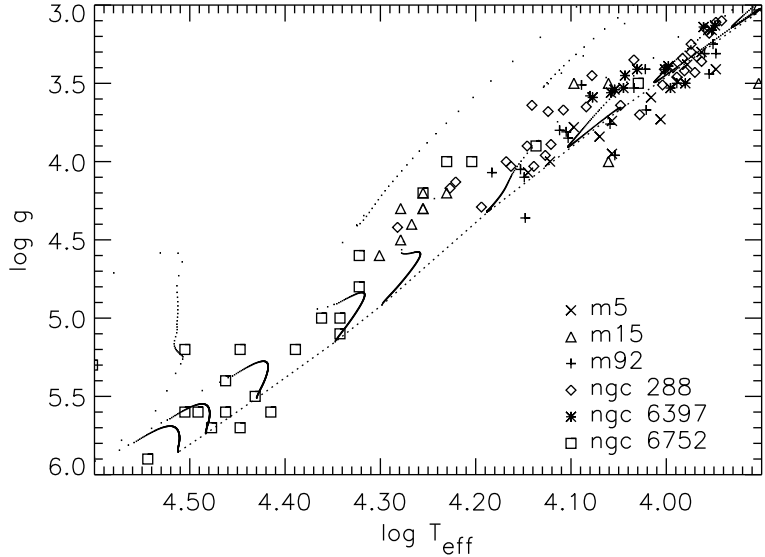}
\caption{ Same as Figure 7 except for helium-mixed tracks with
$\Delta {\rm X}_{\rm mix} \, = \, 0.10$. }
\end{figure}

     The above considerations suggest that helium mixing might be
able to explain the low gravities of the BHB stars.  We explore
this possibility in Figures 8 and 9, where we plot the
observational data from Figure 7 together with the helium-mixed
tracks for $\Delta {\rm X}_{\rm mix} \, = \, 0.10$ and 0.20,
respectively.  The gravities of the tracks in Figure 8 agree well
with the observational data over the temperature range $4.1 \, >
\, {\rm log} \, {\rm T}_{\rm eff} \, > \, 4.0$ but are
still too large for $4.3 \, > \, {\rm log} \, {\rm T}_{\rm
eff} \, > \, 4.2$.  This latter discrepancy can be removed by
more extensive mixing, as shown in Figure 9, but then the
predicted gravities become too low for $4.1 \, > \, {\rm log}
\, {\rm T}_{\rm eff} \, > \, 4.0$.  These results suggest
that the observed variation of log g with ${\rm log} \, {\rm
T}_{\rm eff}$ can be understood if there is a gradient in the
amount of helium mixing, with the hotter BHB stars being more
extensively mixed.  Note that this is in the same sense as is
needed to produce a bluer HB morphology.  For the present
sequences this would imply a variation in ${\rm Y}_{\rm env}$
from $\approx 0.32$ for $4.1 \, > \, {\rm log} \, {\rm
T}_{\rm eff} \, > \, 4.0$ to $\approx 0.44$ for $4.3 \, >
\, {\rm log} \, {\rm T}_{\rm eff} \, > \, 4.2$.  In the
following subsection we will see that an even larger enhancement
of ${\rm Y}_{\rm env}$ may be necessary to explain the EHB
stars.

\begin{figure}[t]
  \figurenum{9}
  \epsscale{0.75}
  \plotone{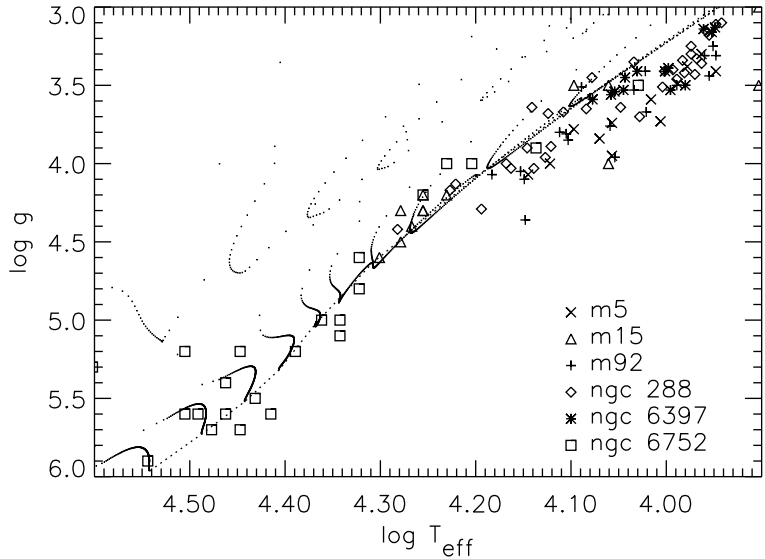}
\caption{ Same as Figure 7 except for helium-mixed tracks with
$\Delta {\rm X}_{\rm mix} \, = \, 0.20$. }
\end{figure}

     The helium-mixed models have lower gravities because they
are more luminous.  This prediction can be tested by comparing
the bolometric luminosities of the helium-mixed models with the
observed luminosities along the BHB, although this test would be
affected by uncertainties in the distance modulus and reddening,
large bolometric corrections to the visual data, small number of
blue standards for photometric calibration, etc.  In addition,
one could use the luminosity difference between the EHB and the
BHB, which is sensitive to the envelope helium abundance, to
differentiate between the canonical and helium-mixed models.  It
would also be important to continue spectroscopic studies of the
BHB stars to look for cases where diffusion has not affected the
surface composition (Crocker, Rood, \& O'Connell 1986; Crocker
\& Rood 1988; Corbally, Gray, \& Philip 1997).

\subsection{Origin of the Extreme HB Stars}
     The origin of the EHB stars found in the field and in some
globular clusters has been a puzzle for some time.  Various
explanations have been offered, including binary-star scenarios
involving mergers (Iben \& Tutukov 1984; Iben 1990) and mass
transfer (Mengel, Norris, \& Gross 1976) as well as single-star
scenarios involving unusually large mass loss along the RGB
(Sweigart, Mengel, \& Demarque 1974; D'Cruz et al. 1996).  The
binary-star scenarios generally predict a wide range in the
luminosity of the EHB stars which appears to be inconsistent with
the observed luminosities of the hot HB stars in the globular
clusters (see, e.g., Landsman et al. 1996b).  For this reason we
will focus our attention on the single-star scenario.

     The principal difficulty of the single-star scenario is the
problem of fine tuning the mass loss along the RGB to produce the
very small envelope masses ($ \approx {\rm 10}^{\rm -2} \,
{\rm M}_{\sun}$) required for the high effective temperatures
($>20,000 \, {\rm K}$) of the EHB stars.  Too much mass loss and a star
will bypass the HB phase and die as a helium white dwarf; too
little mass loss and a star will lie too red along the HB.  We
will now illustrate this problem with some new canonical
calculations.  Following this, we will show how helium mixing may
offer a possible solution because it permits substantially larger
envelope masses.

     Figures 10 and 11 show the tracks for two canonical
sequences during the evolution up the RGB and through the helium
flash to the ZAHB.  The mass loss parameter $\eta_{\rm R}$ for
the sequence in Figure 10 was adjusted to give a ZAHB effective
temperature near the cool end of the EHB.  Due to its high mass
loss rate this sequence was starting to evolve away from the RGB
when the helium flash began.  The gyrations in the track during
the subsequent evolution to the ZAHB are caused by the main and
secondary flashes which occur in the core as the helium burning
progresses towards the center.  This aspect of the helium flash
is discussed more fully by Mengel \& Sweigart (1981) and
Sweigart (1994).

\begin{figure}[t]
  \figurenum{10}
  \epsscale{0.75}
  \protect\plotone{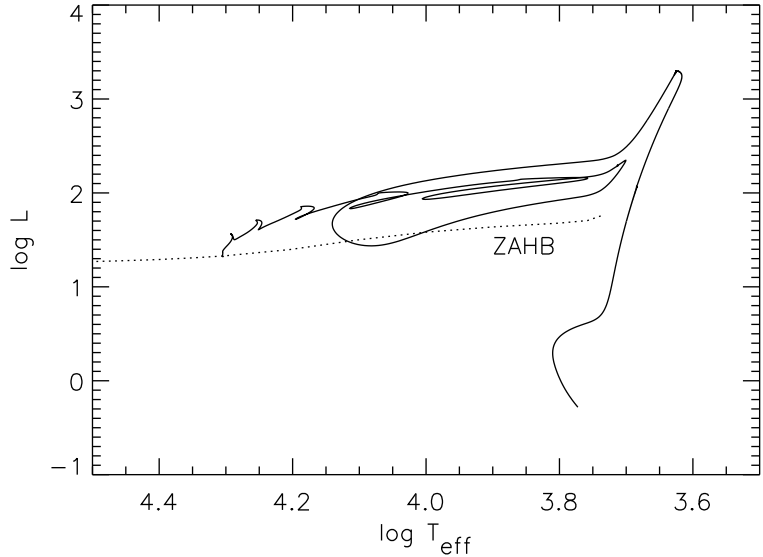}
  \caption{ Canonical evolution from the RGB through the helium
flash to the ZAHB for the Reimers mass loss parameter
$\eta_{\rm R} \, = \, 0.653$. }
\end{figure}

\protect\begin{figure}[t]
  \figurenum{11}
  \epsscale{0.75}
  \protect\plotone{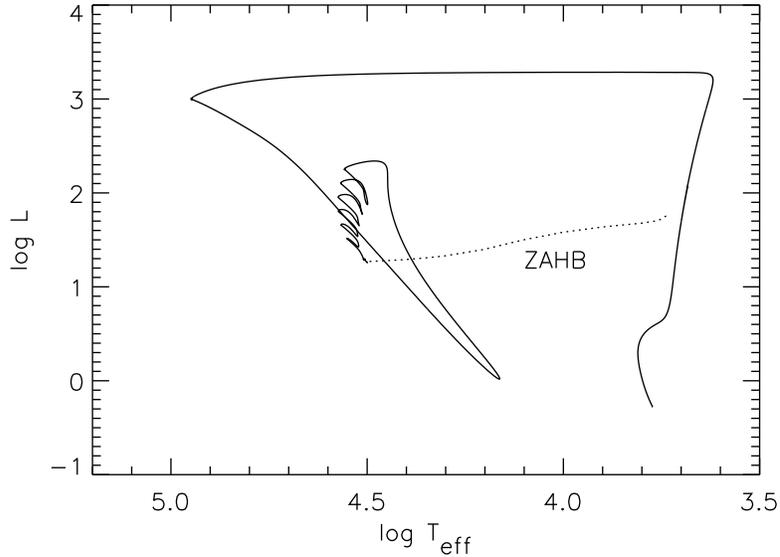}
\caption{\protect Canonical evolution from the RGB through the helium
flash to the ZAHB for the Reimers mass loss parameter
$\eta_{\rm R} \, = \, 0.740$. }
\end{figure}

     The sequence in Figure 11 was computed with a higher mass
loss rate in order to study the hotter EHB stars.  In this case
the models evolved off the RGB to very high effective
temperatures (${\rm log} \, {\rm T}_{\rm eff} \, \approx
\, 5.0$) before undergoing the helium flash.  Following a number
of flash-induced gyrations, this sequence settled onto the hot
end of the EHB.  The ZAHB envelope masses of the sequences in
Figures 10 and 11 were only 0.03 and 0.001 ${\rm M}_{\sun}$,
respectively, as compared to a total mass loss along the RGB of $
\approx 0.3 \, {\rm M}_{\sun}$.  This shows that the ZAHB
envelope mass must fall within a very narrow range compared to
the total mass loss in order to produce an EHB star.

     Castellani \& Castellani (1993) have recently shown that
the helium flash can be further delayed until a star is
descending the white dwarf cooling curve if the mass loss is
sufficiently large.  This possibility has been explored by D'Cruz
et al. (1996), who suggested that the fine tuning problem might
be avoided if the EHB stars were the progeny of such ``hot helium
flashers".  We examine this possibility in Figure 12, where we
follow the evolution of a representative star from the white
dwarf cooling curve to the ZAHB.  The helium flash in this case
is more complicated than in Figures 10 and 11.  At the main flash
peak, the helium-burning luminosity reached ${\rm 10}^{\rm 10}
\, {\rm L}_{\sun}$, as is typical for the helium flash in a
low-mass star.  Ordinarily the flash convection produced by such
high burning rates does not reach the hydrogen shell, and
consequently no hydrogen is mixed from the envelope into the
core.  In contrast, the flash convection in Figure 12 penetrated
into the envelope and rapidly mixed hydrogen into the hot
helium-burning interior.  This hydrogen mixing is analogous to
the mixing that sometimes occurs during a final helium-shell
flash (see Iben 1995 and references therein) and should be a
general characteristic of any helium flash that begins on the
white dwarf cooling curve.

     The penetration of the flash convection into the envelope is
shown more clearly in Figure 13.  At its maximum extent, the
flash convection engulfed nearly 90$\%$ of the envelope
hydrogen, implying a corresponding decrease in the envelope mass
at the ZAHB phase.  Some of the helium and carbon carried out by
the flash convection was subsequently dredged up to the surface
when a convective envelope formed during the rapid redward
excursion of the track following the main flash.  The surface
composition after this dredge-up was 81$\%$ helium and 3$\%$
carbon by mass.  The present calculations do not include the
energy that is released as the hydrogen is mixed inward and
burned, and consequently they may underestimate the surface
composition change.  It is possible that all of the residual
envelope hydrogen may be consumed and that part, or all, of the
carbon may be burned to nitrogen.

     The present results indicate that ``hot helium flashers" are
not the progenitors of the EHB stars both because the predicted
surface abundances seem more appropriate for sdO than sdB stars
and because the predicted ZAHB effective temperatures (${\rm
log} \, {\rm T}_{\rm eff} \, \approx \, 4.6$) are hotter
than those observed in most EHB stars.  The evolution in Figure
12 may, however, be relevant for understanding the origin of the
very helium-rich sdO stars discussed by Lemke et al. (1997).

\begin{figure}[t]
  \figurenum{12}
  \epsscale{0.75}
  \plotone{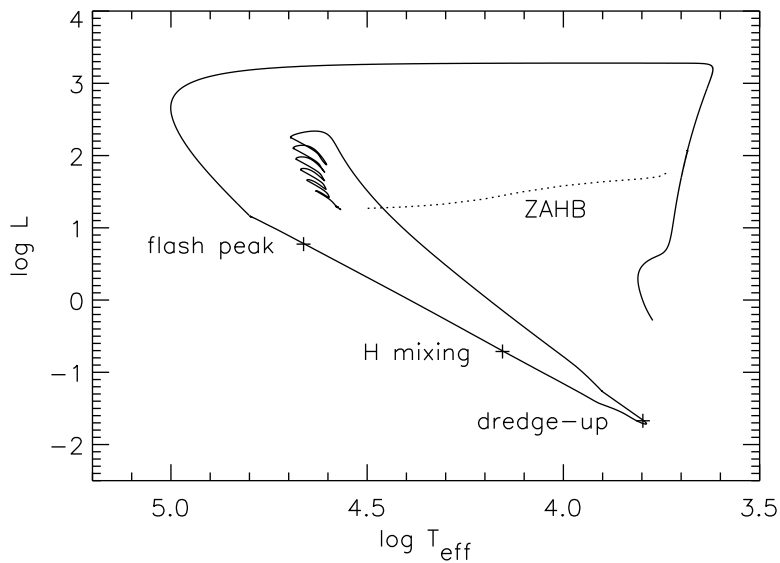}
\caption{ Canonical evolution from the RGB through the helium
flash to the ZAHB for the Reimers mass loss parameter
$\eta_{\rm R} \, = \, 0.750$.  The 3 plus signs denote the
peak of the main helium flash, the onset of hydrogen mixing into
the core and the dredge-up of helium and carbon-rich material by
the convective envelope. }
\end{figure}

     Helium mixing offers an alternative scenario for producing
the EHB stars.  In Sec. 3.1 we saw how substantial helium
mixing could move a star onto the hot end of the HB.  However, we
have not yet determined if helium mixing can overcome the ``fine
tuning" problem discussed previously.  To address this point, we
plot the envelope masses of both the canonical and helium-mixed
models as a function of the ZAHB effective temperature in Figure
14.  We see that EHB stars which have undergone substantial
helium mixing have considerably larger envelope masses than their
canonical counterparts and therefore require much less fine
tuning of the mass loss.  Such models also have a high envelope
helium abundance, which may be important for understanding the
high helium abundances of the sdO progeny of the EHB stars, as we
will discuss in the following subsection.

     There are some observational tests of this helium-mixing
scenario.  Canonical models predict an abrupt increase in the
surface luminosity at the end of the EHB phase as the helium
burning shifts outward from the center to a shell.
Observationally this should appear as a gap between the EHB and
post-EHB stars in either the log L - ${\rm log} \, {\rm
T}_{\rm eff}$ or log g - ${\rm log} \, {\rm T}_{\rm eff}$
diagram.  The helium-mixed models predict a significantly larger
gap, since the hydrogen shell in these models becomes active
during the transition to helium-shell burning as a consequence of
the larger envelope mass.  This effect can be seen by comparing
the tracks in Figure 9 with those in Figure 7.  Although the
statistics are uncertain, there is an indication that the
luminosity gap between the EHB and post-EHB stars in NGC 6752 may
be larger than expected from canonical models (Landsman et al.
1996b).  For these reasons it would be important to study the
separation between the field EHB and post-EHB stars in the log g
- ${\rm log} \, {\rm T}_{\rm eff}$ diagram, where the
statistics are better defined (Saffer 1997).  Such a study might
be complicated, however, by the fact that the abundance
variations in the halo field RGB stars are not as pronounced as
in the globular cluster giants (Suntzeff 1993; Kraft 1994).

\begin{figure}[t]
  \figurenum{13}
  \epsscale{0.75}
  \plotone{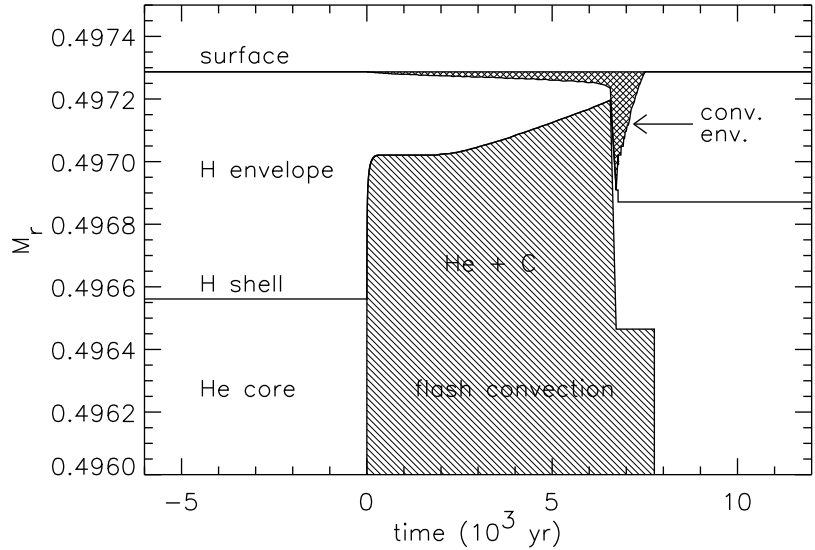}
\caption{ Time dependence of the mass coordinate ${\rm M}_{\rm
r}$ in solar units at the center of the hydrogen shell, the outer
edge of the flash convection zone, the inner edge of the
convective envelope and the stellar surface during the helium
flash shown in Figure 12.  The zero-point of the timescale
corresponds to the peak of the main helium flash.  Shaded areas
are convective. }
\end{figure}

\begin{figure}[t]
  \figurenum{14}
  \epsscale{0.75}
  \plotone{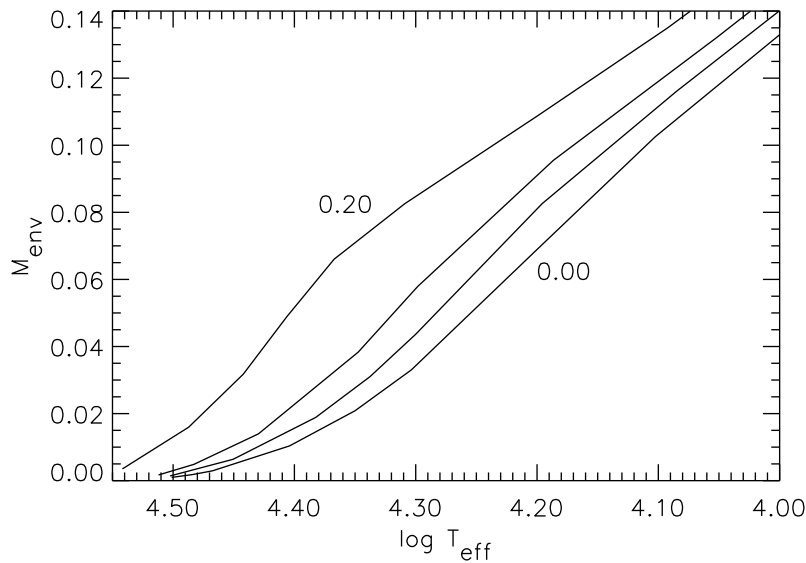}
\caption{ Dependence of the envelope mass ${\rm M}_{\rm env}$
in solar units on ${\rm log} \, {\rm T}_{\rm eff}$ at the
ZAHB phase.  Curves are given for canonical models (labeled 0.00)
and for helium-mixed models with $\Delta {\rm X}_{\rm mix} \,
= \, 0.05$, 0.10 and 0.20 (upper curves). }
\end{figure}

\begin{figure}[t]
  \figurenum{15}
  \epsscale{0.75}
  \plotone{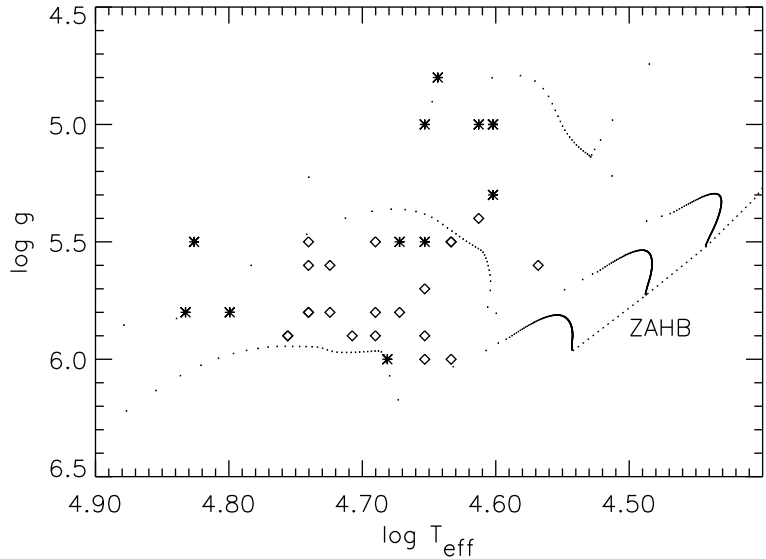}
\caption{ Comparison between the helium-mixed EHB and post-EHB
evolutionary tracks for $\Delta {\rm X}_{\rm mix} \, = \,
0.20$ and the observed gravities of the helium-rich sdO stars.
Data are from Dreizler et al. (1990) (asterisks) and Thejll
(1994) (diamonds). }
\end{figure}

\begin{figure}[t]
  \figurenum{16}
  \epsscale{0.75}
  \plotone{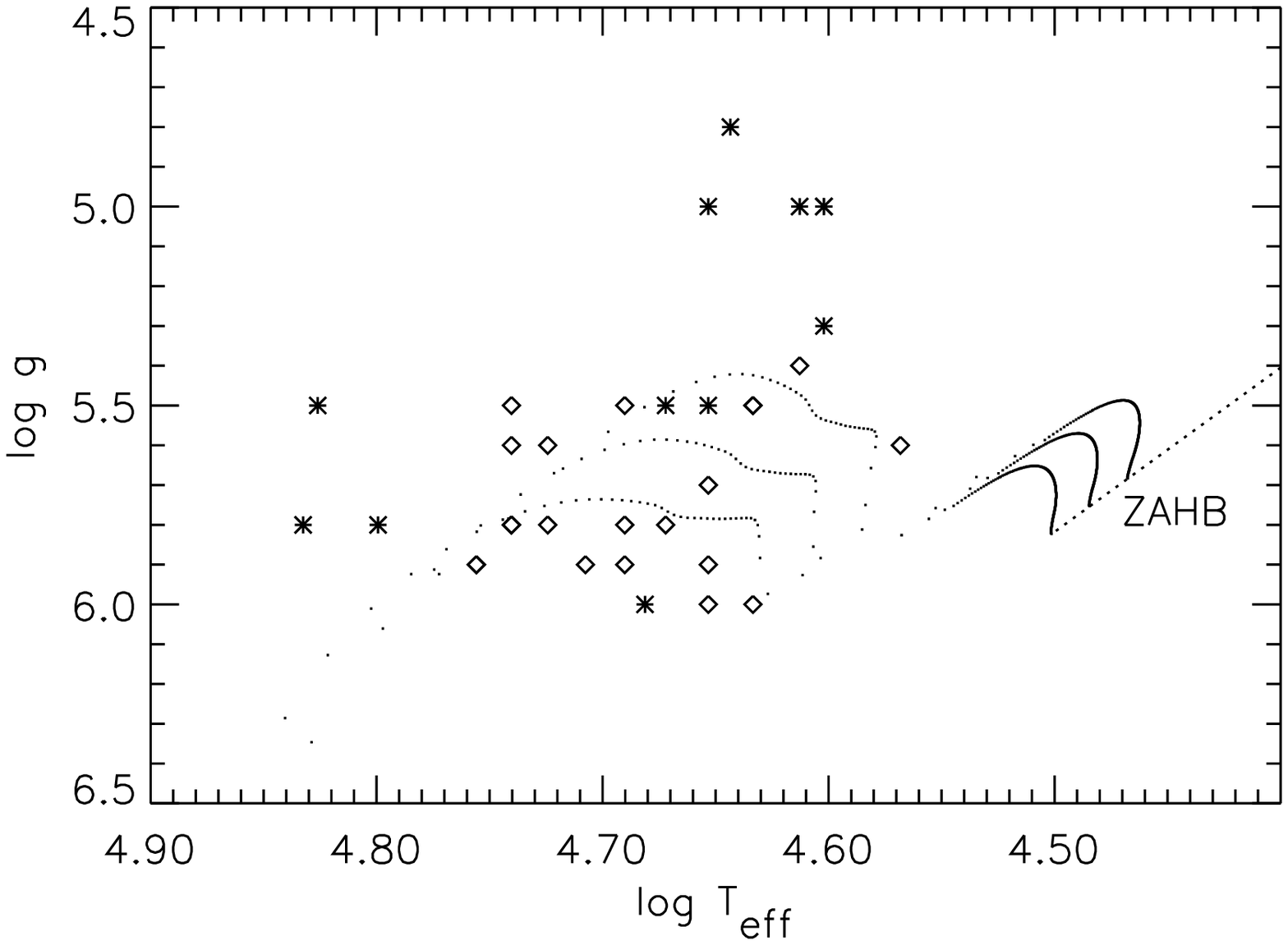}
\caption{ Comparison between canonical EHB and post-EHB
evolutionary tracks and the observed gravities of the helium-rich
sdO stars.  Data are the same as in Figure 15. }
\end{figure}

     One should also search for EHB and post-EHB stars where
diffusion may not have modified the surface composition.  For
example, Moehler, Heber, \& Durrell (1997a) have found a very
hot, helium-rich (${\rm Y} \, \approx \, 0.9$) EHB star in
M15, while Landsman et al. (1996a) have reported on a UV-bright
(and possibly post-EHB) star in $\omega$ Cen with ${\rm Y} \,
\approx \, 0.7$.

\subsection{Evolutionary Status of the sdO Stars}
     A number of evolutionary channels probably contribute to the
field sdO population.  The most luminous, lowest gravity sdO
stars appear to be on post-asymptotic-giant-branch (AGB)
evolutionary tracks, while some of the most helium-rich (and C +
N enhanced) sdO stars may have undergone a delayed helium flash
as in Figure 12.  It is also likely that some sdO stars may have
evolved from the EHB, as suggested by Wesemael et al. (1982).
However, two objections have been raised against this
possibility: 1) that post-EHB evolutionary tracks do not cover
the region of the log g - ${\rm log} \, {\rm T}_{\rm eff}$
diagram occupied by the sdO stars (Thejll 1994) and 2) that the
hydrogen-rich atmosphere of an sdB star cannot be easily
transformed into the helium-rich atmosphere of an sdO star.  We
will now consider whether the helium-mixed models can help to
resolve these objections.

     We address the first objection in Figure 15, where we plot
the 3 EHB and post-EHB evolutionary tracks with the most
extensive helium mixing computed thus far, i.e., $\Delta {\rm
X}_{\rm mix} \, = \, 0.20$, ${\rm Y}_{\rm env} \, \approx
\, 0.6$.  Each track is again represented by a series of dots
separated by an interval of $5 \, \times \, {\rm 10}^{\rm
5}$ yr so that one can see where these models spent the most time
in the log g - ${\rm log} \, {\rm T}_{\rm eff}$ diagram.  The
data for the helium-rich sdO stars in Figure 15 are taken from
Dreizler et al. (1990) and Thejll (1994).  We see that these
helium-rich post-EHB tracks nicely bound the observed location of
the sdO stars, including the sdO stars having the highest surface
gravities.  This suggests that the highest gravity sdO stars do
not necessarily lie on the helium-burning main sequence.

     For comparison we plot 3 canonical EHB and post-EHB tracks
in Figure 16.  These tracks are the hottest ones that we could
compute before the models underwent the extensive flash-induced
mixing shown in Figures 12 and 13.  While the post-EHB tracks in
Figure 16 pass through much of the sdO domain, they cannot
explain the highest gravity sdO stars.  Moreover, these canonical
tracks have very small ZAHB envelope masses ($< 0.003 \, \,
{\rm M}_{\sun}$) and consequently suffer from the fine tuning
problem discussed in the previous subsection.

     The helium-mixed models can straightforwardly overcome the
second objection against the post-EHB scenario for the sdO stars
because the envelopes of these models are already helium-rich.
All that is required is for the outermost layers where helium has
been depleted by diffusion during the sdB phase to be either
removed or mixed with the helium-rich interior.  There are 3 ways
in which this could be accomplished.  First, a hot stellar wind
might remove the thin hydrogen-rich surface veneer.  This
possibility differs from the suggestion of MacDonald \& Arrieta
(1994), who required that the stellar wind remove all of the
envelope in order to expose the helium-rich core.  Secondly,
Groth, Kudritzki, \& Heber (1985) have suggested that HeII/HeIII
convection might mix the outermost layers if ${\rm Y}_{\rm env}
\, > \, 0.5$.  This condition is fulfilled by the helium-mixed
models in Figure 15.  Thirdly, some post-EHB stars experience
helium-shell flashes during which they make brief excursions back
to the AGB (Caloi 1989).  The convective envelope that develops
at that time would mix the envelope and therefore remove any
vestige of the prior diffusion.

     We conclude that the helium-mixed sequences such as those in
Figure 15 support the suggestion of an evolutionary link between
the sdO and sdB stars.

\section{Summary}
     The helium-mixing scenario outlined in this paper predicts
that the effects of helium mixing should become more pronounced
as one goes blueward along the HB.  Thus the RR Lyrae stars
should be affected the least, the BHB stars more so, and the EHB
stars the most.  This trend is in the same sense as was needed to
account for a number of observational results.  Relatively small
amounts of helium mixing can reproduce the 2nd parameter effect
and lower the globular cluster ages derived from the $\Delta
{\rm V}$ method.  Somewhat more mixing is needed to explain the
low gravities of the BHB stars, while extensive mixing seems
necessary to explain the high effective temperatures of the sdB
stars and the high helium abundances of their sdO progeny.
Clearly much observational and theoretical work remains to be
done to explore these implications of helium mixing and to
determine if such mixing actually occurs in low-mass red-giant
stars.

     We conclude with a brief summary of the main points of this
paper:

     1. The observed abundance variations, particularly those
involving Al, suggest that low-mass red-giant stars may be mixing
helium from the top of the hydrogen shell into the envelope.

     2. Noncanonical models with helium mixing have a higher
envelope helium abundance and suffer greater mass loss along the
RGB.

     3. Helium mixing would mimic age as the 2nd parameter and
would reduce the globular cluster age derived from the $\Delta
{\rm V}$ method.

     4. Helium mixing produces a hotter HB morphology, thereby
making it easier to explain the EHB populations observed in
different stellar systems.

     5.  BHB models with helium mixing have lower surface
gravities than canonical models.

     6. Helium mixing might explain the high helium abundances of
some helium-rich sdO stars.

\acknowledgements
The author is indebted to M. Catelan for many insightful
comments, for a critical reading of this paper, and for computing
the HB simulations in Figure 4.  The author also gratefully
acknowledges R. Cavallo for preparing Figure 1.  This research
was supported in part by NASA grant NAG5-3028

\bigskip                                                              
\pagebreak[1]                                                         
\noindent REFERENCES:                                                 
\nopagebreak[4]                                                       
\begin{list}{}{\topsep -\baselineskip                                 
		\partopsep 0in                                        
		\itemsep 0in                                          
		\parsep 0in                                           
		\leftmargin .35in                                     
		\listparindent -.35in                                 
		\labelsep 0in                                         
		\labelwidth 0in}                                      
\item ~                                                               


Arribas, S., Caputo, F., \& Martinez-Roger, C. 1991, A\&AS, 88,
19

Bell, R. A., \& Dickens, R. J. 1980, ApJ, 242, 657

Bell, R. A., Dickens, R. J., \& Gustafsson, B. 1979, ApJ, 229,
604

Bingham, E. A., Cacciari, C., Dickens, R. J., \& Fusi Pecci, F.
1984, MNRAS, 209, 765

Borissova, J., Catelan, M., Spassova, N., \& Sweigart, A. V.
1997, AJ, 113, 692

Briley, M. M., Bell, R. A., Hoban, S., \& Dickens, R. J. 1990,
ApJ, 359, 307

Buonanno, R. 1993, in ASP Conf. Series 48, The Globular Cluster-
Galaxy Connection, ed. G. H. Smith \& J. P. Brodie (San
Francisco: ASP), 131

Buonanno, R., Corsi, C., Bellazzini, M., Ferraro, F. R., \& Fusi
Pecci, F. 1997, AJ, 113, 706

Buonanno, R., Corsi, C. E., \& Fusi Pecci, F. 1985, A\&A, 145,
97

Buonanno, R., Corsi, C. E., \& Fusi Pecci, F. 1989, A\&A, 216,
80

Caloi, V. 1989, A\&A, 221, 27

Carbon, D. F., Langer, G. E., Butler, D., Kraft, R. P., Suntzeff,
N. B., Kemper, E., Trefzger, C. F., \& Romanishin, W. 1982,
ApJS, 49, 207

Castellani, M., \& Castellani, V. 1993, ApJ, 407, 649

Castellani, V., \& De Santis, R. 1994, ApJ, 430, 624

Catelan, M. 1996, private communication

Cavallo, R. M., Sweigart, A. V., \& Bell, R. A. 1996, ApJ, 464,
L79

Cavallo, R. M., Sweigart, A. V., \& Bell, R. A. 1997, in
preparation

Chaboyer, B., Demarque, P., \& Sarajedini, A. 1996, ApJ, 459,
558

Corbally, C. J., Gray, R. O., \& Philip, A. G. D. 1997, this
conference

Crocker, D. A., \& Rood, R. T. 1988, in IAU Sym. 126, Globular
Cluster Systems in Galaxies, ed. J. E. Grindlay \& A. G. D.
Philip (Dordrecht: Kluwer), 509

Crocker, D. A., Rood, R. T., \& O'Connell, R. W. 1986, ApJ, 309,
L23

Crocker, D. A., Rood, R. T., \& O'Connell, R. W. 1988, ApJ, 332,
236

D'Cruz, N. L., Dorman, B., Rood, R. T., \& O'Connell, R. W.
1996, ApJ, 466, 359

de Boer, K. S., Schmidt, J. H. K., \& Heber, U. 1995, A\&A,
303, 95

Denisenkov, P. A., \& Denisenkova, S. N. 1990, Soviet Astron.
Lett., 16, 275

Dorman, B., O'Connell, R. W., \& Rood, R. T. 1995, ApJ, 442, 105

Dreizler, S., Heber, U., Werner, K., Moehler, S., \& de Boer, K.
S. 1990, A\&A, 235, 234

Fusi Pecci, F., Ferraro, F. R., Bellazzini, M., Djorgovski, S.,
Piotto, G., \& Buonanno, R. 1993, AJ, 105, 1145

Fusi Pecci, F., Ferraro, F. R., Crocker, D. A., Rood, R. T., \&
Buonanno, R. 1990, A\&A, 238, 95

Fusi Pecci, F., \& Renzini, A. 1976, A\&A, 46, 447

Fusi Pecci, F., \& Renzini, A. 1978, in IAU Sym. 80, The HR
Diagram, ed. A. G. D. Philip \& D. S. Hayes (Dordrecht: Reidel),
225

Greggio, L., \& Renzini, A. 1990, ApJ, 364, 35

Gross, P. G. 1973, MNRAS, 164, 65

Groth, H. G., Kudritzki, R. P., \& Heber, U. 1985, A\&A, 152,
107

Hill, R. S., et al. 1992, ApJ, 395, L17

Hurley, D. J. C., Richer, H. B., \& Fahlman, G. G. 1989, AJ, 98,
2124

Iben, I., Jr. 1990, ApJ, 353, 215

Iben, I., Jr. 1995, Physics Reports, 250, 1

Iben, I., Jr, \& Renzini, A. 1984, Physics Reports, 105, 329

Iben, I., Jr., \& Tutukov, A. V. 1984, ApJS, 54, 335

Kraft, R. P. 1994, PASP, 106, 553

Kraft, R. P., Sneden, C., Langer, G. E., \& Shetrone, M. D.
1993, AJ, 106, 1490

Kraft, R. P., Sneden, C., Langer, G. E., Shetrone, M. D., \&
Bolte, M. 1995, AJ, 109, 2586

Kraft, R. P., Sneden, C., Smith, G. H., Shetrone, M. D., Langer,
G. E., \& Pilachowski, C. A. 1997, AJ, 113, 279

Landsman, W. B., Crotts, A. P. S., Hubeny, I., Lanz, T.,
O'Connell, R. W., Whitney, J., \& Stecher, T. P. 1996a, in ASP
Conf. Series 99, Cosmic Abundances, ed. S. S. Holt \& G.
Sonneborn (San Francisco: ASP), 199

Landsman, W. B., et al. 1996b, ApJ, 472, L93

Langer, G. E., \& Hoffman, R. D. 1995, PASP, 107, 1177

Langer, G. E., Hoffman, R., \& Sneden, C. 1993, PASP, 105, 301

Lemke, M., Heber, U., Napiwotzki, R., Dreizler, S., \& Engels,
D. 1997, this conference

Liebert, J., Saffer, R. A., \& Green, E. M. 1994, AJ, 107, 1408

MacDonald, J., \& Arrieta, S. S. 1994, in Hot Stars in the
Galactic Halo, ed. S. J. Adelman, A. R. Upgren, \& C. J. Adelman
(Cambridge: Cambridge Univ. Press), 238

Mengel, J. G., Norris, J., \& Gross, P. G. 1976, ApJ, 204, 488

Mengel, J. G., \& Sweigart, A. V. 1981, in IAU Colloq. 68,
Astrophysical Parameters for Globular Clusters, ed. A. G. D.
Philip \& D. S. Hayes (Dordrecht: Reidel), 277

Moehler, S., Heber, U., \& de Boer, K. S. 1995, A\&A, 294, 65

Moehler, S., Heber, U., \& Durrell, P. R. 1997a, A\&A, 317, L83

Moehler, S., Heber, U., \& Rupprecht, G. 1997b, A\&A, in press

Norris, J. 1983, ApJ, 272, 245

Norris, J. 1987, ApJ, 313, L65

Norris, J. E., \& Da Costa, G. S. 1995, ApJ, 441, L81

Paltrinieri, B., et al. 1997, this conference

Peterson, R. C. 1983, ApJ, 275, 737

Peterson, R. C., Rood, R. T., \& Crocker, D. A. 1995, ApJ, 453,
214

Piotto, G., et al. 1997, in Advances in Stellar Evolution, ed. R.
T. Rood \& A. Renzini (Cambridge: Cambridge Univ. Press), in
press

Reimers, D. 1975, Mem. Soc. Roy. Sci. Liege, 6th Series, 8, 369

Renzini, A. 1977, in Advanced Stages in Stellar Evolution, ed. P.
Bouvier \& A. Maeder (Sauverny: Geneva Obs.), 149

Renzini, A. 1983, Mem. S. A. It., 54, 335

Rood, R. T., \& Crocker, D. A. 1989, in IAU Colloq. 111, The Use
of Pulsating Stars in Fundamental Problems of Astronomy, ed. E.
G. Schmidt (Cambridge: Cambridge Univ. Press), 103

Saffer, R. 1997, this conference

Sandage, A. 1982, ApJ, 252, 553

Sandage, A. 1990, ApJ, 350, 631

Sandage, A. 1993a, AJ, 106, 687

Sandage, A. 1993b, AJ, 106, 703

Sandage, A., Katem, B., \& Sandage, M. 1981, ApJS, 46, 41

Silbermann, N. A., \& Smith, H. A. 1995, AJ, 110, 704; erratum:
1996, AJ, 111, 567

Simon, N. R., \& Clement, C. M. 1993, ApJ, 410, 526

Stecher, T. P., et al. 1992, ApJ, 395, L1

Storm, J., Carney, B. W., \& Latham, D. W. 1994, A\&A, 290, 443

Suntzeff, N. 1993, in ASP Conf. Series 48, The Globular Cluster-
Galaxy Connection, ed. G. H. Smith \& J. P. Brodie (San
Francisco: ASP), 167

Sweigart, A. V. 1994, in Hot Stars in the Galactic Halo, ed. S.
J. Adelman, A. R. Upgren, \& C. J. Adelman (Cambridge: Cambridge
Univ. Press), 17

Sweigart, A. V. 1997, ApJ, 474, L23

Sweigart, A. V., \& Mengel, J. G. 1979, ApJ, 229, 624

Sweigart, A. V., Mengel, J. G., \& Demarque, P. 1974, A\&A, 30,
13

Thejll, P. 1994, in Hot Stars in the Galactic Halo, ed. S. J.
Adelman, A. R. Upgren, \& C. J. Adelman (Cambridge: Cambridge
Univ. Press), 197

VandenBerg, D. A., Bolte, M., \& Stetson, P. B. 1996, ARA\&A,
34, 461

VandenBerg, D. A., \& Smith, G. H. 1988, PASP, 100, 314

Wesemael, F., Winget, D. E., Cabot, W., van Horn, H. M., \&
Fontaine, G. 1982, ApJ, 254, 221

Whitney, J. H., et al. 1994, AJ, 108, 1350

\end{list}                                                             

\end{document}